\documentstyle[psfig]{mn}

\def\etal{{\it et al.\ }}
\def\eg{{\it e.g.\ }}

\def\ie{{\it i.e.\ }}

\def\spose#1{\hbox to 0pt{#1\hss}}
\def\approxlt{\mathrel{\spose{\lower 3pt\hbox{$\sim$}}
        \raise 2.0pt\hbox{$<$}}}
\def\approxgt{\mathrel{\spose{\lower 3pt\hbox{$\sim$}}
        \raise 2.0pt\hbox{$>$}}}
\def\approxpropto{\mathrel{\spose{\lower 3pt\hbox{$\sim$}}
        \raise 2.0pt\hbox{$\propto$}}}
\mathchardef\twiddle="2218

\def\multleft#1{\hbox to size{\vbox {\halign {\lft{##}\cr #1}}\hfill}\par}
\def\multright#1{\hbox to size{\vbox {\halign {\rt{##}\cr #1}}\hfill}\par}

\def\today{\ifcase\month\or January\or February\or March\or April\or May\or
      June\or July\or August\or September\or October\or November\or December\fi
      \space\number\day, \number\year}
\def\<{\thinspace}

\font\big=cmr10 scaled\magstep2

\def\erg{{\rm\thinspace erg}}

\def\km{{\rm\thinspace km}}

\def\Mpc{{\rm\thinspace Mpc}}
\def\Msun{\hbox{$\rm\thinspace M_{\odot}$}}

\def\s{{\rm\thinspace s}}

\def\ergps{\hbox{$\erg\s^{-1}\,$}}

\def\kmps{\hbox{$\km\s^{-1}\,$}}

\def\kmpspMpc{\hbox{$\kmps\Mpc^{-1}$}}

\title[Cosmological constraints from the local X-ray luminosity function of  
galaxy clusters]
{Cosmological constraints from the local X-ray luminosity function of the 
most X-ray luminous galaxy clusters}
\author[S.W. Allen et al.]
{\parbox[]{6.in} {S.W. Allen$^1$, R.W. Schmidt$^1$, A.C. Fabian$^1$ and H. Ebeling$^2$ \\
\footnotesize
1. Institute of Astronomy, Madingley Road, Cambridge CB3 0HA \\
2. Institute for Astronomy, 2680 Woodlawn Drive, Honolulu, Hawaii 96822, USA
 }}
\begin{document}
\maketitle
\begin{abstract}
We present precise constraints on the normalization of the power
spectrum of mass fluctuations in the nearby universe, $\sigma_8$,  as
a function of the mean local matter density, $\Omega_{\rm m}$. Using
the observed local X-ray luminosity function of galaxy clusters from
the extended BCS and REFLEX studies, a mass-luminosity relation
determined from Chandra and ROSAT X-ray data and weak gravitational
lensing observations, and the mass function predicted by the Hubble
Volume simulations of Evrard \etal,  we obtain $\sigma_8 =
(0.508\pm0.019) \,\Omega_{\rm m}^{-(0.253\pm0.024)}$,  with
$\Omega_{\rm m} < 0.34$ at 68 per cent confidence.  The degeneracy
between $\sigma_8$ and $\Omega_{\rm m}$ can be broken using Chandra
measurements of the X-ray gas mass fractions in dynamically relaxed
clusters. Using this information and including Gaussian priors on the
mean baryon density of the universe and the Hubble constant,  we
obtain $\sigma_8=0.695\pm0.042$ and $\Omega_{\rm m}=0.287\pm0.036$, for
an assumed flat $\Lambda$CDM cosmology (marginalized 68 per cent
confidence limits). Our results are in good agreement with some recent
studies based on the local X-ray temperature function of clusters, the
redshift evolution of the X-ray luminosity and temperature functions
of clusters, early results from the Sloan Digitized Sky Survey, the
most recent results from studies of cosmic shear, and combined analyses
of the 2dF galaxy redshift survey and cosmic microwave background
anisotropies.
\end{abstract}


\section{Introduction}

The X-ray luminosity function of galaxy clusters in the nearby
universe  provides a powerful cosmological probe. The observed
luminosity function,  $n(L)$, can be combined with a relation linking
the observed X-ray luminosity and mass, and the mass function, $n(M)$,
predicted by simulations, to obtain  tight constraints on the
combination of cosmological parameters $\Omega_{\rm m}$ and
$\sigma_8$, where $\Omega_{\rm m}$ is the mean matter density of the
local universe and $\sigma_8$ is the root-mean-square (rms)
variation of the density field smoothed by a top hat window function
of size 8$h^{-1}$Mpc.

Observationally, the keys to such studies are precise determinations
of the local X-ray luminosity function of clusters and the relation
linking the observed X-ray luminosities and total masses. X-ray
selection currently offers the best way to identify massive galaxy
clusters, and the local X-ray luminosity function has now been precisely
determined by the BCS (Ebeling \etal 1997; Ebeling \etal 2000)  and
REFLEX (B\"ohringer \etal 2002) studies. The flux-limited BCS and
REFLEX  samples, which  are based on data from the ROSAT All-Sky
Survey (RASS; Tr\"umper 1993), together include $\sim 750$ clusters
and cover approximately two thirds of the sky.  Recently, significant
effort has also been invested into measuring the local temperature
function of clusters (\eg Markevitch 1998; Pierpaoli \etal 2001; Ikebe
\etal 2002), which offers a complementary  method for determining
cosmological parameters. At present, however, the temperature function
samples are significantly smaller than the combined BCS-plus-REFLEX
luminosity function data set, and the analysis of the temperature
function is complicated by the fact that these samples are selected
according to X-ray flux as well as temperature, requiring the use
of both mass-temperature and temperature-luminosity relations in the 
analysis.

Recent years have also seen significant efforts directed towards a
precise calibration of the `virial' relations linking the observed
luminosities, temperatures and masses of galaxy clusters (\eg Horner,
Mushotzky \& Scharf 1999; Nevalainen, Markevitch \&  Forman 2000;
Finoguenov, Reiprich  \& B\"ohringer 2001; Allen, Schmidt \&  Fabian
2001b; Sanderson \etal 2003). In particular, the launch of the
Chandra X-ray Observatory has permitted the first precise measurements
of the temperature and mass profiles of relaxed clusters from X-ray
data.  Using a combination of Chandra  and gravitational lensing data,
Allen \etal  (2001b) confirmed that luminous, relaxed galaxy clusters
follow the simple scaling relations predicted by theory, but that the
normalization  of the observed mass-temperature relation measured
within $r_{2500}$ (where the mean enclosed mass density is 2500 times
the critical density of the universe at the redshifts of the clusters)
is approximately 40 per cent lower than predicted by standard
adiabatic simulations. This highlights the likely importance of
additional physics such as cooling and pre-heating in the intracluster
gas (see also  Pearce \etal 2000; Thomas \etal 2002; Voit \etal 2002;
Muanwong \etal 2002).

Theoretically, the primary requirement for cosmological studies using
the observed luminosity and/or temperature functions  of clusters is a
precise prediction of the mass function. This has now been achieved
for flat $\Lambda$CDM (and $\tau$CDM) cosmologies using the Hubble
Volume simulations  of Jenkins \etal (2001) and Evrard \etal (2002).

In this paper we present precise constraints on $\sigma_8$  and
$\Omega_{\rm m}$ based on the observed local luminosity  function of
the most X-ray luminous clusters in the extended BCS  (Ebeling \etal
2000) and REFLEX samples, and a new calibration, using pointed Chandra
and ROSAT X-ray observations and weak gravitational lensing results,
of the  mass-luminosity  relation linking the masses of clusters
measured within $r_{200}$ to their total $0.1-2.4$ keV ROSAT
luminosities.  Having determined our combined constraint on $\sigma_8$
and $\Omega_{\rm m}$, we show that the degeneracy between these
parameters can be broken using Chandra results on the X-ray gas mass
fractions in the most dynamically relaxed clusters. Including Gaussian
priors  on the  mean baryon density of the universe ($\Omega_{\rm
b}h^{2} = 0.0205\pm0.0018$; O'Meara \etal 2001), the Hubble constant
($h=0.72\pm0.08$; Freedman \etal 2001),  and a theoretical bias factor
($b=0.93\pm0.05$; Bialek, Evrard \& Mohr 2001) relating the asymptotic
baryon fraction in the most X-ray luminous clusters to the mean value for
the universe as a whole, we obtain  $\sigma_8=0.695\pm0.042$ and
$\Omega_{\rm m}=0.287\pm0.036$ (marginalized 68 per cent confidence
limits for an assumed flat $\Lambda$CDM cosmology). We compare our
results to other measurements based on the local number density of
clusters, evolution of the X-ray luminosity and temperature functions,
the 2dF galaxy redshift survey, cosmic microwave background
anisotropies, and measurements of cosmic shear.

Throughout this paper, a flat $\Lambda$CDM cosmology with
$\Omega_{\Lambda} = 1 - \Omega_{\rm m}$ is assumed.  In order to
facilitate a direct comparison with previous X-ray studies, results on
the masses, X-ray luminosities and X-ray gas mass fractions of individual
clusters are quoted for a Hubble parameter $h =H_0/100$\kmpspMpc $=
0.5$ or, equivalently, $h_{50} =H_0/50$\kmpspMpc $= 1.0$.

\section{Theory: The predicted luminosity function of galaxy clusters}\label{section:theory}

Jenkins \etal (2001) show that the predicted mass function of galaxy
clusters of mass $M$ at redshift $z$ can be written as a function of
$\ln \sigma^{-1}(M,z)$, where $\sigma^2(M,z)$ is the variance of the
linearly evolved density field smoothed by a spherical top-hat filter
of comoving radius  $R$, enclosing a mass $M=4 \pi R^3
\bar{\rho}_0/3$. Here $\bar{\rho}_0=\Omega_{\rm m}(0)\rho_{\rm c}(0)$
is the mean comoving matter density of the universe, $\Omega_{\rm
m}$(0) is the mean, present matter density in units of the critical
density, and $\rho_{\rm c}(0)={3 H_0^2}/{8 \pi G}$  is the critical
density at redshift zero.

Using a spherical overdensity  algorithm to measure the masses of
clusters  within radii $r_{200}$, where the mean enclosed mass density
is 200 times the critical density of the universe at the redshift of
interest, Evrard  \etal (2002) show that for a flat $\Lambda$CDM 
cosmology, the mass fraction $f(\sigma^{-1})$ can be written as

\vspace{0.3cm}
\begin{equation} 
f(\sigma^{-1}) \ = \ A \exp\big[-|\ln\sigma^{-1}+ B
 \,|^{\epsilon}\big] \,,
\end{equation}

\vspace{0.3cm}

\noindent where, for $\Omega_{\rm m}=0.3$ and $z=0$, $A=0.22$, 
$B=0.73$ and $\epsilon=3.86$. 
Evrard \etal (2002) also provide simple interpolations for $A, B$ and
$\epsilon$ for other values of $\Omega_{\rm m}$ and $z$. The differential
number density of clusters with mass $M$ at redshift $z$ is

\vspace{0.3cm}
\begin{equation}
{{\rm d\,}n(M,z) \over {\rm d\,ln} \,\sigma^{-1}} = \frac{f(\sigma^{-1}) \bar{\rho}(z)}{M},
\end{equation}

\vspace{0.3cm}

\noindent where $\bar{\rho}(z)=\bar{\rho}_0(1+z)^3$ is the mean mass density 
of the universe at redshift $z$. Following Viana \& Liddle (1999), we write
 
\vspace{0.3cm}
\begin{equation}
\label{sig}
\sigma(R,z)=\sigma_{8}(z)\left(\frac{R}{8 h^{-1}\;{\rm 
Mpc}}\right)^{-\gamma(R)},\,
\end{equation}

\noindent where
 
\begin{equation}
\label{del}
\gamma(R)=(0.3\Gamma+0.2)\left[2.92+\log_{10}
        \left(\frac{R}{8 h^{-1}\;{\rm Mpc}}\right)\right]\,
\end{equation}

\vspace{0.3cm}
\noindent and $\Gamma$ is the shape parameter of the cold dark
matter transfer function. Following Sugiyama (1995), we set 

\vspace{0.3cm}
\begin{equation}
\label{del}
\Gamma = \Omega_{\rm m}(0) h \left( \frac{2.7 {\rm K}}{T_0} \right)^2  \exp \left( -\Omega_{\rm b}(0) - \sqrt{\frac{h}{0.5}}\frac{\Omega_{\rm b}(0)}{\Omega_{\rm m}(0)} \right), \,
\end{equation} 

\vspace{0.3cm}
\noindent where $T_0=2.726$K is the temperature of the cosmic
microwave background (Mather \etal 1994) and $\Omega_{\rm b}(0) =
(0.0205\pm0.0018) h^{-2}$  is the mean, present-day
baryon density in units of the critical density (O'Meara \etal 2001). 
For this calculation, we set $h=0.72$ and fix the primordial
spectral index $n=1$. The redshift-dependent quantity $\sigma_{8}(z)$ 
is related to its present value, $\sigma_8(0)$, by

\vspace{0.3cm}
\begin{equation}
\label{growth}
\sigma_8(z) = \sigma_8(0) \, \frac{g(\Omega_{\rm m}(z))}{g(\Omega_{\rm m}(0))} \, 
\frac{1}{1+z} \,,
\end{equation}

\vspace{0.3cm}
\noindent where, for a flat $\Lambda$CDM universe

\vspace{0.3cm}
\begin{equation}
g(\Omega_{\rm m}(z)) = \frac{5}{2} \Omega_{\rm m}(z) \left[ \frac{1}{70} +
        \frac{209\Omega_{\rm m}(z)}{140} -\frac{{\Omega_{\rm m}(z)}^2}{140} + {\Omega_{\rm m}(z)}^{4/7}
        \right]^{-1} \,
\end{equation}

\noindent and 

\vspace{0.09cm}
\begin{equation}
\Omega_{\rm m}(z)  =  \Omega_{\rm m}(0) \, \frac{(1+z)^3}{1-\Omega_{\rm m}(0) +(1+z)^3 \Omega_{\rm m}(0)}.
        \quad \quad 
\end{equation}

\vspace{0.3cm}

\noindent \eg Viana \& Liddle (1996). Combining equations 2 and 3, we obtain 

\vspace{0.2cm}
\begin{equation}
{{\rm d\,}n(M,z) \over {\rm d\,}M}  = \frac{ \gamma \bar{\rho}(z)}{3 M^2}
f(\sigma^{-1}).
\end{equation}

\vspace{0.3cm}

\noindent In Section~\ref{section:ml} we show that for clusters with 
X-ray luminosities, $L$, exceeding $10^{45}\,h_{50}^{-2}$ \ergps~in
the $0.1-2.4$  keV ROSAT band, the mass $M$ measured within $r_{200}$
can be related  to the X-ray luminosity by a power-law model of the form

\vspace{0.3cm}
\begin{equation}
E(z) M = M_0 { \left[ \frac{L}
{E(z)} \right] }^\alpha,
\end{equation} 

\vspace{0.3cm}

\noindent where the best fit values, uncertainties on $M_0$ and 
$\alpha$, and the intrinsic scatter in the relation, are determined 
from Chandra and ROSAT X-ray data and weak gravitational lensing results.
The evolution parameter

\vspace{0.3cm}
\begin{equation}
E(z) =(1+z)\sqrt{(1+z\Omega_{\rm
m}+\Omega_{\Lambda}/(1+z)^2-\Omega_{\Lambda})} 
\end{equation}

\vspace{0.3cm}

\noindent accounts for the expected cosmological scaling of the 
relationship (\eg Bryan \& Norman 1998; Mathiesen \& Evrard 2001). 
Applying the chain rule to equations 9 and 10, we obtain 

\vspace{0.3cm}
\begin{equation}
{{\rm d\,}n(L,z) \over {\rm d\,}L}  = \frac{\gamma \bar{\rho}(z) \alpha }{3 M_0} \left[\frac{E(z)}{L} \right]^{\alpha+1} f(\sigma^{-1}).
\end{equation}

\vspace{0.3cm}
\noindent The predicted differential luminosity function  (\ie the
comoving number density of clusters at redshift  $z$ with luminosities
in an interval ${\rm d}L$ around $L$) given by equation 12 can be
compared with the observed X-ray luminosity function from the BCS and
REFLEX studies to constrain the combination of cosmological
parameters $\Omega_{\rm m}(0)$ and $\sigma_8(0)$, hereafter referred
to as  $\Omega_{\rm m}$ and $\sigma_8$, respectively.

\section{Observations}

\subsection{The observed X-ray luminosity function of the most luminous 
galaxy clusters in the RASS}\label{section:bcs}

For this study, we concentrate on the local ($z \approxlt 0.3$) X-ray
luminosity function and restrict ourselves to the most luminous
clusters, with  $L_{\rm X} > 10^{45}\,h_{50}^{-2}$\ergps~in the
$0.1-2.4$ keV ROSAT band (for a flat $\Lambda$CDM cosmology with
$\Omega_{\rm m}=0.3$ and $\Omega_\Lambda=0.7$). This selection is
facilitated by the large sample size and well-determined selection
functions of the BCS and REFLEX data sets.

The restriction to high luminosities reduces systematic uncertainties
by matching the luminosity range of the luminosity function data to
the range over which the mass-luminosity relation has been calibrated
(Section~\ref{section:ml}).  At lower luminosities, the effects of
pre-heating and cooling in  the intracluster gas are expected to
become important and cause the mass-luminosity and mass-temperature
relations to deviate from simple power-law forms (\eg Cavaliere, Menci
\& Tozzi 1997). Since the  most massive clusters provide the most
powerful constraints on cosmological  parameters, relatively little
information is lost by restricting ourselves to the largest
systems. 

The binned X-ray luminosity functions for  clusters with $L_{\rm
X,0.1-2.4} > 10^{45}\,h_{50}^{-2}$\ergps~from the northern extended
BCS (Ebeling \etal 2000) and  southern REFLEX (B\"ohringer \etal 2002)
studies are summarized in Table~\ref{table:lumfunc}. The mean redshift
of the BCS clusters in this luminosity range is $z=0.21$.\footnote{We
make the comparison between the observed and predicted luminosity
functions at $z=0.21$, the mean redshift of the BCS clusters with
$L_{\rm X,0.1-2.4} > 10^{45}\,h_{50}^{-2}$\ergps. Shifting this
redshift by $\pm0.05$ does not significantly change the results.}

\begin{table}
\begin{center}
\caption{The observed, binned X-ray luminosity function of the
most X-ray luminous ($L_{\rm X,0.1-2.4} > 10^{45}\,h_{50}^{-2}$\ergps)
galaxy clusters from the extended BCS and REFLEX studies. Column 2
gives the mean $0.1-2.4$ keV X-ray luminosity for each bin, $L_{\rm
X}$, in  $10^{44}\,h_{50}^{-2}$\ergps. Error  bars indicate the bin
boundaries. Column 3 gives the number of clusters in each bin and
column 4 the comoving space density in
$h_{50}^5$\,Mpc\,$^{-3}$\,($10^{44}$\ergps)$^{-1}$.   A $\Lambda$CDM
cosmology with $\Omega_{\rm m}=0.3$ and $\Omega_\Lambda=0.7$ is
assumed.}\label{table:lumfunc} 
\vskip -0.1truein
\begin{tabular}{ c c c c c }
&&&& \\
                 & ~ & $L_{\rm X}$                   & $n_{\rm clus}$ &   $n(L)$                      \\
\hline
BCS              & ~ &  $11.73^{+1.62}_{-1.73}$      & 17             & $1.32\pm0.32\times10^{-9}$  \\
                 & ~ &  $15.65^{+2.20}_{-2.30}$      & 17             & $7.45\pm1.81\times10^{-10}$ \\
                 & ~ &  $23.91^{+28.8}_{-6.06}$      & 17             & $8.10\pm1.97\times10^{-11}$ \\
                              
REFLEX           & ~ &  $11.25^{+2.19}_{-1.19}$      & 20             & $1.56\pm0.34\times10^{-9}$ \\
                 & ~ &  $16.27^{+4.61}_{-2.83}$      & 20             & $4.44\pm0.98\times10^{-10}$ \\
                 & ~ &  $29.95^{+76.2}_{-9.07}$      & 20             & $1.77\pm0.39\times10^{-11}$ \\
\hline                      
\end{tabular}
\end{center}
\end{table}

\subsection{The observed mass-luminosity relation for the most X-ray luminous 
galaxy clusters}

In determining the mass-luminosity relation we have used mass
measurements obtained from Chandra X-ray observations of dynamically
relaxed clusters and weak gravitational lensing results drawn from the
literature. X-ray luminosities are determined from pointed ROSAT
observations. In total, the sample  used to define the mass-luminosity
relation includes 17 clusters with precise mass estimates and X-ray
luminosities $L_{X,0.1-2.4}> 10^{45}\,h_{50}^{-2}$\ergps, spanning the
redshift range $0.08<z<0.47$.

\subsubsection{Chandra mass measurements}

\begin{table}
\begin{center}
\caption{Summary of the Chandra observations. The dates of the
observations are given in column 3.  Column 4 lists the exposure times
in ks.}\label{table:chandra_obs}
\vskip -0.1truein
\begin{tabular}{ c c c c }
                 &      z  &    Date  & Exposure      \\  \hline Abell
478        &   0.088 &    2001 Jan 27 & 42.4    \\  PKS0745-191      &
0.103 &    2001 Jun 16 & 17.9    \\  Abell 963        &   0.206 &
2000 Oct 11 & 36.3    \\  Abell 2390       &   0.230 &    1999 Nov 07
& 9.1     \\  Abell 2667       &   0.233 &    2001 Jun 19 & 9.6     \\
Abell 1835       &   0.252 &    1999 Dec 12 & 19.6    \\  Abell 611
&   0.288 &    2001 Nov 03 & 36.1    \\  MS2137-2353      &   0.313 &
1999 Nov 18 & 20.6    \\ RXJ1347-1145     &   0.451 &    2000 Mar
05/Apr 29 & 18.9 \\  3C295            &   0.461 &    1999 Aug 30 &
17.0    \\  &&& \\ \hline
\end{tabular}
\end{center}
\end{table}

\begin{table*}
\begin{center}
\caption{Summary of the Chandra mass measurements.  Column 2 gives the
evolution parameter, $E(z)$,  appropriate for each cluster. Columns 3
and 4 summarize the best-fitting NFW model parameters: the scale
radius, $r_{s}$ (in $h_{50}^{-1}$\,Mpc) and concentration parameter,
$c$. Columns 5 and 6 give the virial  radii, $r_{200}$ (in
$h_{50}^{-1}$\,Mpc) and  masses, $M_{200}$ (in
$10^{14}\,h_{50}^{-1}$\,\Msun). Error bars are 68 per cent confidence
limits for a single interesting parameter, determined from the $\chi^2$
grids.  A flat $\Lambda$CDM cosmology with  $\Omega_{\rm m}=0.3$ and
$\Omega_{\rm \Lambda}=0.7$ is assumed.}\label{table:nfw} 
\vskip 0.05truein
\begin{tabular}{c c c c  c c c }
\multicolumn{1}{c}{} &
\multicolumn{1}{c}{} &
\multicolumn{1}{c}{} & 
\multicolumn{1}{c}{} &
\multicolumn{1}{c}{} &
\multicolumn{1}{c}{} &
\multicolumn{1}{c}{} \\              
              & ~ & $E(z)$ &  $r_{s}$                &       c                  &   $r_{200}$             &       $M_{200}$           \\       
\hline                                                                                                                               
Abell 478     & ~ & 1.042   & $0.94^{+0.18}_{-0.11}$  & $3.67^{+0.31}_{-0.35}$  &  $3.45^{+0.27}_{-0.17}$ &  $25.8^{+6.7 }_{-3.4 }$   \\
PKS0745-191   & ~ & 1.050   & $0.90^{+0.13}_{-0.17}$  & $3.83^{+0.52}_{-0.27}$  &  $3.44^{+0.20}_{-0.27}$ &  $26.0^{+4.9 }_{-5.6 }$   \\
Abell 963     & ~ & 1.107   & $0.42^{+0.14}_{-0.07}$  & $5.72^{+0.78}_{-1.07}$  &  $2.40^{+0.20}_{-0.15}$ &  $9.86^{+2.74}_{-1.77}$   \\
Abell 2390    & ~ & 1.122   & $1.06^{+2.23}_{-0.55}$  & $3.20^{+1.79}_{-1.57}$  &  $3.40^{+1.95}_{-0.82}$ &  $28.9^{+83.6}_{-16.2}$   \\
Abell 2667    & ~ & 1.124   & $0.98^{+0.67}_{-0.29}$  & $3.02^{+0.74}_{-0.85}$  &  $2.96^{+0.63}_{-0.38}$ &  $19.0^{+14.9}_{-6.4 }$   \\
Abell 1835    & ~ & 1.135   & $0.77^{+0.25}_{-0.13}$  & $4.21^{+0.53}_{-0.61}$  &  $3.24^{+0.44}_{-0.19}$ &  $25.5^{+11.7}_{-4.2 }$   \\
Abell 611     & ~ & 1.158   & $0.56^{+0.97}_{-0.25}$  & $4.58^{+2.36}_{-2.22}$  &  $2.56^{+1.04}_{-0.43}$ &  $13.1^{+23.3}_{-5.5 }$   \\
MS2137-2353   & ~ & 1.174   & $0.22^{+0.04}_{-0.04}$  & $8.71^{+1.22}_{-0.92}$  &  $1.95^{+0.12}_{-0.14}$ &  $5.95^{+1.17}_{-1.23}$   \\
RXJ1347-1145  & ~ & 1.271   & $0.52^{+0.25}_{-0.17}$  & $6.34^{+1.61}_{-1.35}$  &  $3.28^{+0.56}_{-0.50}$ &  $33.2^{+19.9}_{-13.0}$   \\
3C295         & ~ & 1.279   & $0.22^{+0.10}_{-0.06}$  & $7.90^{+1.71}_{-1.72}$  &  $1.77^{+0.22}_{-0.18}$ &  $5.27^{+2.23}_{-1.43}$   \\
&&&&& \\
\hline
\end{tabular}
\end{center}
 \parbox {7in}
{}
\end{table*}

The Advanced CCD Imaging Spectrometer (ACIS) on Chandra permits
direct, simultaneous measurements of the X-ray gas temperature  and
density profiles and, via the hydrostatic  assumption, the total mass
distributions in galaxy clusters.  We have used Chandra to obtain
precise mass measurements for a sample of ten of the most X-ray
luminous, dynamically relaxed clusters  identified from the RASS. The
relaxed dynamical states of the clusters are demonstrated by their
regular X-ray and optical morphologies, X-ray temperature maps and, in
6/10 cases, from the availability of  consistent, independent mass
measurements from gravitational  lensing studies
(see Section~\ref{section:systematics}).

The Chandra observations were made using the ACIS and the
back-illuminated S3 detector between 1999  August 30 and 2001 November
3.  We have used the level-2 event lists provided by the standard
Chandra pipeline processing. These lists were cleaned for periods of
background flaring using the CIAO software package, resulting in the
net exposure times summarized in Table~\ref{table:chandra_obs}.

The data have been analysed using the methods described by Allen \etal
(2001a, 2002b) and Schmidt \etal (2001; these papers present detailed
mass analyses of Abell 2390, RXJ1347-1145 and Abell 1835,
respectively).  In brief, concentric annular spectra were extracted
from the cleaned event lists,  centred on the peaks of the X-ray
emission from the clusters. (For  RXJ1347-1145, the data from the
southeast quadrant of the cluster were  excluded due to ongoing merger
activity in that region; Allen \etal 2002b.) The spectra were analysed
using XSPEC (version 11.0: Arnaud 1996), the MEKAL plasma emission
code (Kaastra \& Mewe 1993; incorporating the Fe-L  calculations of
Liedhal, Osterheld \& Goldstein 1995), and the photoelectric
absorption  models of Balucinska-Church \& McCammon (1992; the
absorbing column density  was included as a free parameter in the
fits, alleviating  problems associated with uncertainties in
the quantum efficiency of the  detectors at low energies). Only data
in the $0.5-7.0$ keV energy range were used. The spectra for all
annuli were modelled simultaneously in order to determine the
deprojected X-ray gas temperature profiles, under the assumption of
spherical symmetry.

\begin{table}
\begin{center}
\caption{Summary of the mass results based on the Dahle \etal (2002) weak
lensing observations. Columns 2 and 3 summarize the redshifts and
evolution parameters for the clusters. Column 4 lists the virial
masses, $M_{200}$ (in $10^{14}\,h_{50}^{-1}$\,\Msun), determined from 
fits to the observed tangential shear profiles using NFW models with a fixed
concentration parameter, $c=5$. No correction for the effects of 
correlated substructure has been applied. Error bars are 68 per 
cent confidence limits for a single interesting parameter.  
A $\Lambda$CDM cosmology
with $\Omega_{\rm m}=0.3$ and $\Omega_{\rm \Lambda}=0.7$ is
assumed. }\label{table:dahle} 
\vskip -0.1truein
\begin{tabular}{ c c c c c }
&&&& \\
            & ~ & $z$    & $E(z)$ &   $M_{200}$           \\
\hline                                                            
Abell 520   & ~ & 0.203  & 1.106  & $14.3^{+5.0}_{-4.3}$  \\ 
Abell 209   & ~ & 0.206  & 1.107  & $ 4.5^{+2.4}_{-2.0}$  \\
Abell 963   & ~ & 0.206  & 1.107  & $ 7.0^{+3.1}_{-2.6}$  \\
Abell 141   & ~ & 0.230  & 1.122  & $13.6^{+7.9}_{-6.2}$  \\
Abell 267   & ~ & 0.230  & 1.122  & $15.5^{+4.4}_{-3.9}$  \\
Abell 1576  & ~ & 0.299  & 1.165  & $17.5^{+4.3}_{-3.8}$  \\ 
Abell 1995  & ~ & 0.320  & 1.179  & $20.1^{+4.6}_{-4.2}$  \\ 
Abell 1351  & ~ & 0.328  & 1.184  & $42.3^{+7.8}_{-6.9}$  \\
\hline                      
\end{tabular}
\end{center}
\end{table}

For the mass modelling, azimuthally averaged surface brightness
profiles were constructed from background subtracted, flat-fielded
images with a $0.984\times0.984$ arcsec$^2$ pixel scale ($2\times2$
raw detector pixels). When combined with the deprojected spectral
temperature profiles,  the surface brightness profiles can be used to
determine the X-ray gas mass and total
mass profiles in the clusters.\footnote{The observed surface
brightness profile and a particular parameterized mass model are
together used to predict the temperature profile of the X-ray gas. (We
use the median temperature profile determined from 100 Monte-Carlo
simulations. The outermost pressure is fixed using an iterative
technique which ensures a smooth pressure gradient in these regions.)
The predicted temperature profile is rebinned to the same binning as
the spectral results and the $\chi^2$  difference between the observed
and predicted, deprojected temperature  profiles is calculated. The
parameters for the mass model are  stepped through a regular grid of
values in the $r_{\rm s}$-$c$ or $r_{\rm s}$-$\sigma$  planes to 
determine the best-fitting values and 68 per cent confidence
limits. (The best-fit models generally provide good descriptions of
the data). The gas mass profile is determined to high precision at
each grid point directly from the observed surface  brightness profile
and model temperature profile. Spherical symmetry and  hydrostatic
equilibrium are assumed throughout. Small departures from spherical 
symmetry are expected to have a negligible effect on the 
results \eg Piffaretti \etal (2003). } For this analysis, we have used
an enhanced version of the image deprojection code described by White,
Jones \& Forman (1997) with distances calculated using the code of
Kayser, Helbig \& Schramm (1997).

We have parameterized the cluster  mass (luminous plus dark matter)
profiles using a  Navarro, Frenk \& White (1997; hereafter NFW) model
with

\begin{equation}
\rho(r) = {{\rho_{\rm c}(z) \delta_{\rm c}} \over {  ({r/r_{\rm s}}) 
\left(1+{r/r_{\rm s}} \right)^2}},
\end{equation}


\noindent where $\rho(r)$ is the mass density,  $\rho_{\rm c}(z) =
3H(z)^2/ 8 \pi G$ is the critical density for closure at redshift $z$,
$r_{\rm s}$ is the scale  radius, $c$ is the concentration parameter
($c=r_{200}/r_{\rm s}$) and  $\delta_{\rm c} = {200 c^3 / 3 \left[
{{\rm ln}(1+c)-{c/(1+c)}}\right]}$. The normalizations of the mass
profiles can also be expressed in terms of an effective velocity
dispersion, $\sigma = \sqrt{50}\, r_{\rm s} c H(z)$  (with $r_{\rm s}$
in units of Mpc and $H(z)$ in \kmpspMpc).

The best-fit NFW parameter values and 68 per cent confidence limits
are summarized in Table~\ref{table:nfw}. This table also lists  the
`virial' radii, $r_{200}$, where the mean enclosed  density is 200
times the critical density of the universe at the  redshifts of the
clusters, and the masses within these radii, $M_{200}$. (The
uncertainties on parameters are determined directly from the 
$\chi^2$ grids.) Note that the Chandra data only cover the
central  regions of the clusters out to radii $0.2-0.5\, r_{200}$ and
thus some extrapolation of the models, assuming that the NFW
parameterization remains valid to $r_{200}$, is required in
calculating the virial masses.

\subsubsection{Weak lensing mass measurements}\label{section:dahle}

In order to expand the sample of clusters used to construct the
mass-luminosity relation, and allow certain tests of the fairness of
this relation (Section~\ref{section:systematics}), we have also
included data for clusters with precise mass measurements from wide
field weak gravitational lensing studies.  In particular, we have
included data from the study of Dahle \etal (2002), who present
aperture mass profiles for eight clusters with
$L_{X,0.1-2.4}>10^{45}\,h_{50}^{-2}$\ergps~
(Section~\ref{section:luminosities}) obtained from wide-field imaging
with the University of Hawaii (UH) 2.24m telescope and UH8K
camera. One of these clusters, Abell 963, is also in our Chandra
sample. Since the Chandra and lensing data for this cluster  give
consistent $M_{200}$ results, but the Chandra data provide  tighter
constraints, we use the Chandra result in our  default analysis. (The
Dahle \etal 2002 mass measurement  for Abell 963  is, however, used in
our analysis of the weak lensing subsample, discussed in
Section~\ref{section:systematics}.)  The weak lensing mass
measurements made with the UH8K camera used a control annulus of 550
arcsec, which corresponds to 3.4\,$h_{50}^{-1}$Mpc for a cluster at
$z=0.3$.

Unlike X-ray mass measurements, which are based on the hydrostatic
assumption, lensing mass measurements are independent of the dynamical
state of the  gravitating matter. As a result,  there are no
restrictions in the Dahle  \etal (2002) sample on the dynamical
states of the clusters; several of the systems appear to be undergoing
major merger events. The inclusion of the Dahle \etal (2002) clusters
allows us both to examine the effects of dynamical activity on the
mass-luminosity relation and  assess whether the use of Chandra data
for dynamically relaxed clusters  is likely to bias our determination
of cosmological parameters (see  Section~\ref{section:systematics}).

We have used the aperture mass profiles presented by Dahle \etal
(2002) to recover the mean tangential shear profiles for the clusters
and have fitted these with NFW models. In general, the lensing data
are unable to constrain both  the concentration parameter and scale
radius of the NFW models and so we have fixed $c=5$ for this analysis,
a typical value for such massive clusters inferred from simulations
(\eg Navarro \etal 1997), and consistent with the Chandra results
listed in Table~\ref{table:nfw}. The masses of the clusters determined
from the Dahle \etal (2002) data are summarized  in
Table~\ref{table:dahle}.

In addition to the Dahle \etal (2002) UH8K data, accurate  weak
lensing mass measurements are also available for Abell 2390 (Squires
\etal 1996) and  RXJ1347-1145 (Fischer \& Tyson 1997).  In both cases
the lensing mass measurements at $r_{200}$ are in good agreement with
the Chandra results (Allen \etal 2001a, 2002b). Dahle \etal (2002)
also present a weak lensing mass measurement for Abell 1835 using a
smaller camera, which is consistent with the Chandra result in
Table~\ref{table:nfw}.

On the basis of numerical simulations, Metzler, White \& Loken (2001)
argue that large scale structure in the environments ($r \sim 10-20
h^{-1}$ Mpc) of galaxy clusters are likely cause measurements of
$M_{200}$ from weak lensing to overestimate the true masses of
clusters by, on average, $\sim 30$ per cent. Previous work by Cen
(1997) and Reblinksy \& Bartelmann (1999) had argued for smaller
effects, of the order of $\sim 10$ per cent. Based on these studies,
we have included a statistical correction of 20 per cent to the
lensing masses in our determination of  cosmological parameters \ie
we multiply the masses in Table~\ref{table:dahle} by 0.83. We note
that the  effects of more distant, uncorrelated structure along the
lines of sight to the clusters are not expected to bias the lensing
mass measurements  (\eg Metzler \etal 2001, Hoekstra 2002).

\begin{table*}
\begin{center}
\caption{Summary of the pointed ROSAT observations. Columns 2 and 3
list the date of observation and the detector used. Column 4 gives
the exposure times in ks. Column 5 lists the intrinsic $0.1-2.4$
luminosities, $L_{0.1-2.4}$ (in $10^{44}\,h_{50}^{-2}$\ergps). Error 
bars are 68 per cent confidence limits. A $\Lambda$CDM cosmology with
$\Omega_{\rm m}=0.3$ and $\Omega_{\rm \Lambda}=0.7$ is assumed.}
\label{table:rosat}
\vskip 0.2truein
\begin{tabular}{ c c c c c c c c c c c }
             &  Date & Detector & Exposure   & $L_{0.1-2.4}$ \\
\hline                                        
Abell 478    &  1991 Aug 31 & PSPC & 21.4    & $22.7\pm0.2$ \\
PKS0745-191  &  1993 Oct 15 & PSPC & 10.5    & $28.2\pm0.6$ \\
Abell 520    &  1998 Mar 09 & HRI  & 27.7    & $18.0\pm0.8$ \\    
Abell 209    &  1996 Jul 01 & HRI  & 10.6    & $15.2\pm1.0$ \\    
Abell 963    &  1991 Apr 20 & HRI  & 10.5    & $13.4\pm1.0$ \\
Abell 141    &  1996 Dec 10 & HRI  & 16.2    & $12.6\pm0.7$ \\    
Abell 267    &  1996 Jan 03 & HRI  & 15.7    & $11.1\pm0.9$ \\    
Abell 2390   &  1993 Nov 13 & PSPC & 10.3    & $31.7\pm0.5$ \\
Abell 2667   &  1994 Dec 14 & HRI  & 21.3    & $29.2\pm1.1$ \\
Abell 1835   &  1993 Jul 03 & PSPC & 6.2     & $38.3\pm0.9$ \\
Abell 611    &  1996 Apr 04 & HRI  & 17.3    & $11.4\pm1.1$ \\
Abell 1576   &  1993 Nov 07 & PSPC & 16.3    & $13.4\pm0.4$ \\    
MS2137-2353  &  1993 Nov 07 & PSPC & 10.5    & $19.1\pm0.9$ \\
Abell 1995   &  1995 Nov 13 & HRI  & 16.5    & $13.7\pm1.4$ \\    
Abell 1351   &  1995 Apr 29 & HRI  & 31.9    & $16.5\pm3.0$ \\ 
RXJ1347-1145 &  1995 Jan 28 & HRI  & 15.8    & $81.1\pm3.1$ \\
3C295        &  1995 Jun 19 & HRI  & 22.2    & $12.2\pm1.3$ \\
&&&& \\                                       
\hline                                                
\end{tabular}
 \end{center}
\end{table*}

\begin{figure}
\vspace{0.5cm} \hbox{
\hspace{-0.0cm}\psfig{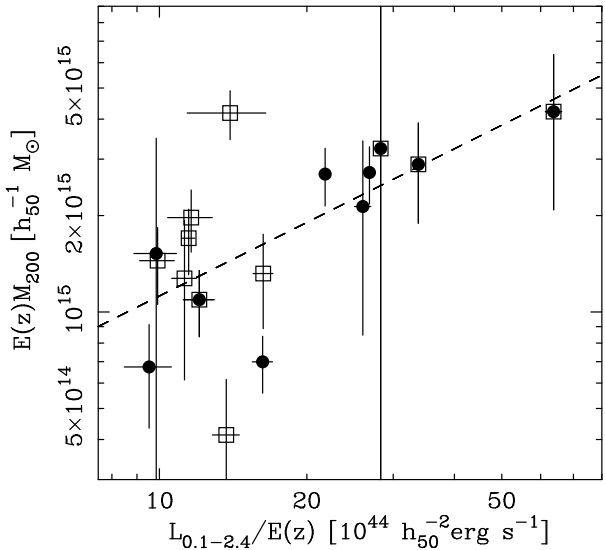}
} \caption{The observed mass-luminosity relation. Chandra mass
measurements for dynamically relaxed clusters  are indicated by filled
circles. Weak lensing mass measurements are indicated by open
squares. A 20 per cent correction for the effects of correlated
substructure has been applied to the lensing masses
(Section~\ref{section:dahle}). The four clusters with Chandra mass
measurements and consistent weak  lensing results are indicated by
filled circles surrounded by open squares. The data for Abell 267 have
been offset slightly for display purposes. The best-fitting power law
model from Section~\ref{section:ml} is shown as the dashed curve. The
two most significant outliers above (Abell 1351) and below (Abell 209)
the best-fit curve appear to be undergoing  major merger events (see
Section~\ref{section:systematics}).  A $\Lambda$CDM cosmology with
$\Omega_{\rm m}=0.3$ and $\Omega_{\rm \Lambda}=0.7$  is
assumed.}\label{fig:ml}
\end{figure}

\subsubsection{ROSAT luminosity measurements}\label{section:luminosities}

In constructing the mass-luminosity relation, we have used pointed
ROSAT observations to determine the total, intrinsic $0.1-2.4$ keV
luminosities of the clusters. This minimizes systematic uncertainties
by matching the observing band and (as far as possible) detector
technology to that used for the RASS observations. The agreement
between the BCS and REFLEX fluxes,  which are based on RASS data, and
the fluxes determined from deep, pointed ROSAT observations is good
(Ebeling \etal 1998; B\"ohringer \etal 2002).  The details of the
pointed ROSAT observations are summarized in
Table~\ref{table:rosat}. The data were analysed using the XSELECT
package (version 2.0) and XSPEC (version 11.0).  The emission-weighted
temperatures and metallicities of  the clusters were set to the values
measured with Chandra. Where Chandra data were not available, the
metallicity was set to 0.3 solar and the temperature was determined
iteratively from the luminosity-temperature relation of Allen \&
Fabian  (1998). Note, however, that the precise settings of the
temperatures and metallicities  have little effect on the measured
$0.1-2.4$ keV luminosities for such hot, massive clusters. The
absorbing column densities were set to the Galactic values determined
by  Dickey \& Lockman (1990).

The intrinsic $0.1-2.4$ keV luminosities for a $\Lambda$CDM cosmology 
with  $\Omega_{\rm m}=0.3, \Omega_{\rm \Lambda}=0.7$ and $h=0.5$ 
(the same cosmology assumed in the BCS and REFLEX luminosity functions in 
Table~\ref{table:lumfunc}) are listed in Table~\ref{table:rosat}.

\subsubsection{The observed mass-luminosity relation}\label{section:ml}

\begin{figure}
\vspace{0.5cm}
\hbox{
\hspace{-0.5cm}\psfig{figure=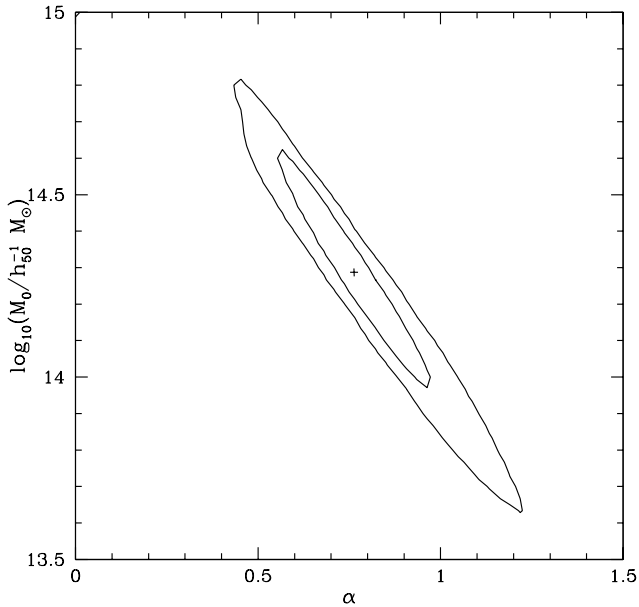,width=.49\textwidth,angle=0}
} \caption{The 68.3 and 95.4 per cent confidence contours on the 
intercept and slope of the model fitted to the 
mass-luminosity data. The plus sign marks the best-fit position. 
}\label{fig:m0_alpha}
\end{figure}

The mass-luminosity relation for the 17 X-ray luminous clusters in
our sample, measured within radii $r_{200}$ corresponding to a density
contrast $\Delta=200$ with respect to the critical density of the
universe at the redshifts of the  clusters, is shown in
Fig.~\ref{fig:ml}. Those clusters with mass measurements from Chandra
X-ray data are indicated by filled circles. Clusters with weak lensing
mass measurements are marked with open squares. The four clusters with
Chandra mass measurements and consistent weak lensing results are
indicated  by filled circles surrounded by open squares. A
flat $\Lambda$CDM cosmology with  $\Omega_{\rm m}=0.3$ and 
$\Omega_{\rm \Lambda}=0.7$ is assumed.

We have fitted the data using a model of the form


\begin{eqnarray}
{\rm log_{10}} \left[ \frac{E(z)\,M_{200}}{h_{50}^{-1}\Msun} \right]~ = ~\alpha\, { {\rm log_{10}} \left[ \frac{L}{E(z) \,10^{44}\,h_{50}^{-2}{\rm \ergps} } \right] } \nonumber\\
+~\,{\rm log_{10}}\left[\frac{M_0}{h_{50}^{-1}\Msun}\right]. \hspace{2.0cm}
\end{eqnarray}

\noindent Using the BCES($Y|X$) estimator of Akritas \& Bershady
(1996), which  accounts for errors in both axes and the presence of
possible intrinsic scatter, we obtain best-fitting values and 68 per
cent confidence limits from $10^6$ bootstrap simulations of
log$_{10}(M_0/h_{50}^{-1}\,M_{\odot})=14.29^{+0.20}_{-0.23}$
and $\alpha=0.76^{+0.16}_{-0.13}$. The distribution of $M_0$ and $\alpha$
values are shown in Fig.~\ref{fig:m0_alpha}.  The observed slope is in
good agreement with the  expected slope of the mass--{\it bolometric}
luminosity relation of $\alpha=0.75$, from models of simple gravitational 
collapse.

The rms scatter of the observed log\,[$E(z)M_{200}$] values about  the
best fitting curve is $0.22$ for the full sample of 17 clusters, 
0.15 for the 10 dynamically relaxed clusters studied with Chandra, 
and $0.29$ for the 8 clusters with weak lensing measurements 
from Dahle \etal (2002). The scatter in
the full sample is similar to that measured by Reiprich \&
B\"ohringer (2001) using ASCA and ROSAT X-ray data.  Note, however,
that our best-fitting curve implies a slightly  higher X-ray
luminosity for a given mass.

\section{Determination of cosmological parameters}

\subsection{Monte Carlo method}

Using the theoretical prescription for the mass function of galaxy
clusters  described in Section~\ref{section:theory}, the observed
BCS and REFLEX X-ray luminosity functions summarized in Table 1, 
and the mass-luminosity data discussed in Section~\ref{section:ml}, 
we can determine $\sigma_8$ as a function of $\Omega_{\rm m}$.

In determining our results on cosmological parameters, we have used a
Monte Carlo approach. For each iteration of the code, we construct a
random bootstrap sample of the $M_{200}$ and $L_{0.1-2.4}$ values
listed in Tables $3-5$ and examine a grid  of ($96\times131$)
$\Omega_{\rm m}$ and $\sigma_8$ values, covering the plane  $0.05<
\Omega_{\rm m} <1.0$ and $0.20 < \sigma_8 <1.50$.  For each value of
$\Omega_{\rm m}$, we scale the mass and luminosity values
appropriately and determine the best-fitting mass-luminosity
relation. For each $\Omega_{\rm m}$, $\sigma_8$ parameter pair, we
construct a model X-ray luminosity function, including the effects of
random scatter in the mass-luminosity relation, which we characterize
using a log-normal distribution.  The model X-ray luminosity function
is then compared with the observed BCS and REFLEX  data, also scaled
to the appropriate cosmology, and the $\chi^2$  difference between the
two is calculated. In this way, the best fitting  $\Omega_{\rm m}$,
$\sigma_8$ pair for the grid is determined, which provides us with a
single sample result. The whole process is repeated for $10^6$
iterations to produce the final results, discussed below.

\subsection{Results on $\sigma_8$ as a function of $\Omega_{\rm m}$}

\begin{figure}
\vspace{0.5cm}
\hbox{
\hspace{0.2cm}\psfig{figure=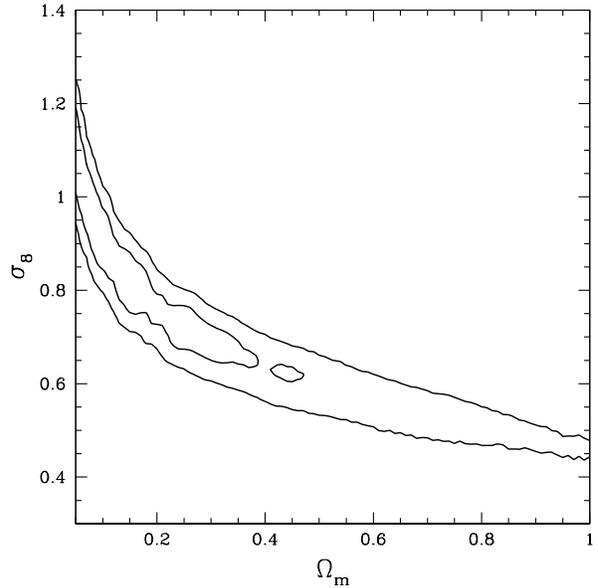,width=.46\textwidth,angle=0}
} \caption{The 68.3 and 95.4 per cent confidence contours 
on $\sigma_8$ and $\Omega_{\rm m}$ obtained from the BCS+REFLEX 
luminosity function data (Table~\ref{table:lumfunc}), theoretical
mass function, and observed mass-luminosity relation
(Section~\ref{section:ml}). A flat $\Lambda$CDM cosmology is
assumed.}\label{fig:contour1}
\end{figure}

Fig.~\ref{fig:contour1} shows the 68.3 and 95.4 per cent  confidence
contours in the $\sigma_8-\Omega_{\rm m}$ plane from one million
iterations of the Monte Carlo code. The results  exhibit the well
known degeneracy between $\sigma_8$ and $\Omega_{\rm m}$, which can be
approximated (for $0.1<\Omega_{\rm m}<0.4$) by the simple fitting
formula $\sigma_8 = (0.510\pm0.019) \,\Omega_{\rm
m}^{-(0.253\pm0.024)}$.  Our analysis favours low values for
$\Omega_{\rm m}$: marginalizing over $\sigma_8$, we find $\Omega_{\rm
m} < 0.34$ at  68 per cent confidence.

Note that when neglecting the effects of scatter and uncertainties in the 
normalization and slope of the mass-luminosity relation, we obtain
the best fit ($\chi^2=5.1$ for 4 degrees of freedom) for 
$\Omega_{\rm m}=0.23$ and $\sigma_8=0.74$.

\subsection{Breaking the $\sigma_8 - \Omega_{\rm m}$ degeneracy using the 
Chandra $f_{\rm gas}(z)$ data}

The usual approach, discussed in the literature, to break the
degeneracy between $\sigma_8$ and $\Omega_{\rm m}$ 
(Fig.~\ref{fig:contour1}) is to use the redshift evolution of the
luminosity and/or temperature function of clusters, which depends
strongly on the mean  mass density of the universe (see  references in
Section~\ref{section:evolution}). However, this approach is both
observationally challenging in terms of identifying complete,
high-redshift cluster samples, and prone to systematic uncertainties
due to potentially increased levels of dynamical activity and
contaminating AGN emission at high redshifts, which introduce
additional scatter  into the mass-luminosity and mass-temperature
relations.

Fortunately, the Chandra data offer a powerful alternative method to
break the degeneracy between $\sigma_8$ and $\Omega_{\rm m}$ using the
observed X-ray gas mass fractions, $f_{\rm gas}$, in the clusters and
their apparent redshift dependence (\eg White \& Frenk 1991; White
\etal 1993; Sasaki 1996; Pen 1997; Ettori \& Fabian 1999;  Allen \etal
2002a; Erdogdu, Ettori \& Lahav 2002; Ettori, Tozzi \& Rosati
2003). The matter content of rich clusters of galaxies is thought to
provide an almost fair sample of the matter content of the universe
(\eg White \etal 1993). The observed ratio of baryonic to total mass
in clusters is therefore expected to closely match  the ratio of the
cosmological parameters  $\Omega_{\rm b}/\Omega_{\rm m}$, where
$\Omega_{\rm b}$ is the mean baryon density of the universe. The
apparent redshift dependence of the $f_{\rm gas}$ measurements arises
from the fact that the measured $f_{\rm gas}$ values depend upon the
assumed angular diameter distances  to the sources as $f_{\rm gas}
\propto D_{\rm A}^{1.5}$. Thus, although we expect the measured
$f_{\rm gas}$ values to be invariant  with redshift, this will only
appear to be the case when the assumed cosmology  matches the true,
underlying cosmology.

\begin{figure}
\vspace{0.5cm}
\hbox{
\hspace{-0.5cm}\psfig{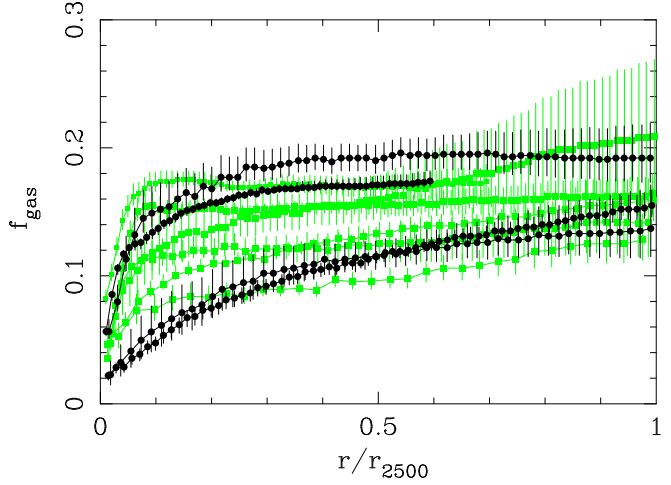} }
\caption{The X-ray gas mass fraction profiles measured with Chandra, 
with the radial  axes scaled in units of $r_{2500}$.  The results for the  six
clusters previously discussed by Allen \etal (2002a) are shown in
lighter  shading. The new results for Abell 478, 611, 963 and 2667 are 
shown as dark circles. Note that $f_{\rm gas}(r)$ is an integrated
quantity and so the  error bars on neighbouring points in a profile
are correlated. An SCDM cosmology with $h=0.5$ is assumed.}\label{fig:fgas_r}
\end{figure}

The observed $f_{\rm gas}$ profiles for the ten relaxed clusters
studied with Chandra, for an assumed $h=0.5$ standard cold dark matter (SCDM)
cosmology with $\Omega_{\rm m}=1.0$ and  $\Omega_{\rm \Lambda}=0.0$,
are shown in Fig.~\ref{fig:fgas_r}. (The six clusters previously
studied by Allen \etal 2002a are shown in a lighter shading.) With the
possible exception of Abell 963, we see that the $f_{\rm gas}$
profiles appear to have converged, or be close to converging, within
$r_{2500}$. We note that the data for the two nearest clusters, Abell 478
and PKS0745-191, do not extend to $r_{2500}$. However, their $f_{\rm gas}$ 
profiles appear to be close to converging within $r \sim 0.6 r_{2500}$. 

\begin{figure}
\vspace{0.5cm}
\hbox{
\hspace{-0.0cm}\psfig{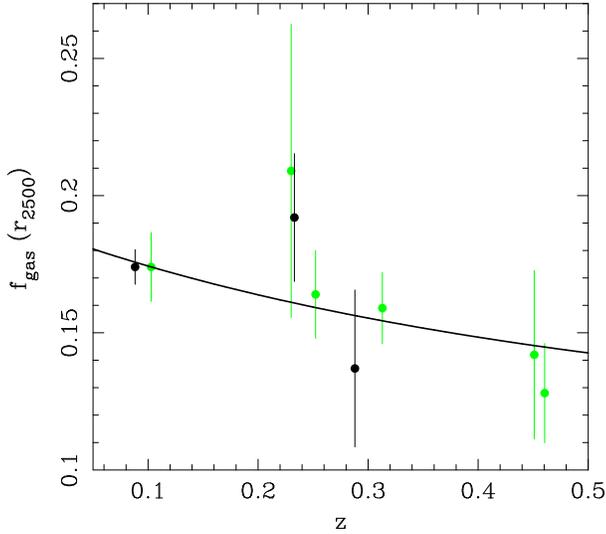}}
\caption{The apparent redshift variation of the X-ray gas mass
fraction measured at $r_{2500}$ (with root-mean-square $1\sigma$
errors) for the nine clusters with convergent $f_{\rm gas}$ profiles
in Fig.~\ref{fig:fgas_r} (see text). The  results for the  six
clusters previously discussed by Allen \etal (2002a)  are shown in
lighter shading.  The solid curve shows the predicted  $f_{\rm
gas}(z)$ behaviour for a flat $\Lambda$CDM cosmology with $\Omega_{\rm
m}=0.291$.}\label{fig:fgas_z}
\end{figure}

\begin{figure}
\vspace{0.5cm} \hbox{
\hspace{-0.5cm}\psfig{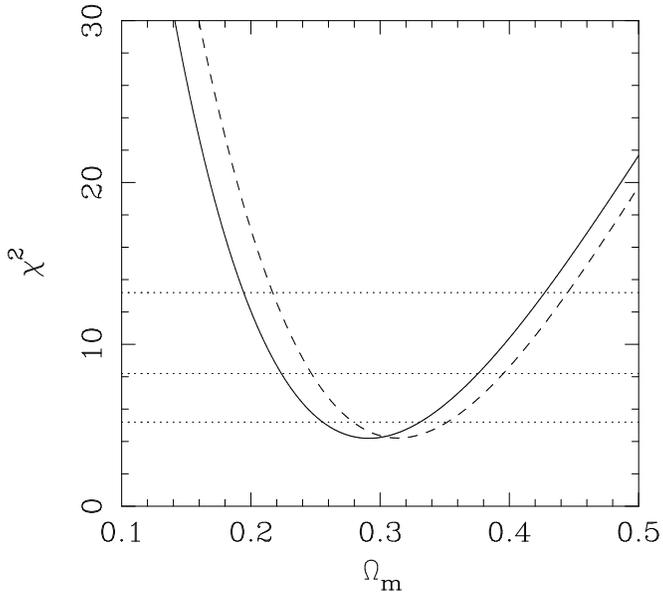}
} \caption{The constraints on $\Omega_{\rm m}$  from the Chandra
$f_{\rm gas}(z)$ data in Fig.~\ref{fig:fgas_z}. A flat  $\Lambda$CDM
cosmology and Gaussian priors of $h=0.72\pm0.08$ and  $\Omega_{\rm
b}h^{2} =  0.0205\pm0.0018$ are assumed. The solid curve shows the
results obtained using a Gaussian prior on the bias factor,
$b=0.93\pm0.05$. The dashed curves show the results for $b=1$ (fixed).
The 1, 2 and 3 sigma confidence limits are marked as dotted
lines.}\label{fig:fgas_chi}
\end{figure}

Fig.~\ref{fig:fgas_z} shows the $f_{\rm gas}$ measurements at
$r_{2500}$ (or at the outermost radii studied for Abell
478 and PKS0745-191) as a function of redshift, for the nine clusters
with convergent  $f_{\rm gas}$ profiles.  Following Allen \etal
(2002a), we have fitted these data with the model

\begin{eqnarray}
f_{\rm gas}^{\rm mod}(z) = \frac{ b\, \Omega_{\rm b}} {\left(1+0.19
\sqrt{h}\right) \Omega_{\rm m}} \left[ \frac{h}{0.5} \, \frac{D_{\rm
A}^{\Omega_{\rm m}=1, \Omega_{\rm  \Lambda}=0 }(z)}{D_{\rm
A}^{\Omega_{\rm m},\,\Omega_{\Lambda}=1-\Omega_{\rm m}}(z)}
\right]^{1.5}
\end{eqnarray}

\noindent and determined the best-fitting value of $\Omega_{\rm m}$, for
an assumed flat $\Lambda$CDM cosmology. The parameter $b$ is a bias
factor that is motivated by gasdynamical simulations, which suggest
that the baryon fraction in clusters is slightly depressed with
respect to the universe as a whole (\eg Cen \& Ostriker 1994; Eke,
Navarro \& Frenk 1998a; Frenk \etal 1999; Bialek \etal 
2001). We include a Gaussian prior on the bias factor,
$b=0.93\pm0.05$, a value appropriate for hot ($kT>5$ keV), massive
clusters in the redshift  range $0<z<0.5$ from the simulations of
Bialek \etal (2001). We also include Gaussian priors on the Hubble
constant, $h=0.72\pm0.08$, the final result from the Hubble Key
Project reported by Freedman \etal (2001), and $\Omega_{\rm b}h^{2} =
0.0205\pm0.0018$ (O'Meara \etal 2001), from cosmic nucleosynthesis
calculations constrained by the observed abundances of light  elements
at high redshifts. The constraints on $\Omega_{\rm m}$ determined from
this analysis are shown as the dark, solid curve in
Fig.~\ref{fig:fgas_chi}.  We obtain $\Omega_{\rm
m}=0.291^{+0.040}_{-0.036}$ at 68 per cent  confidence. Also shown
(dashed curve) are the results obtained when fixing  the bias
parameter $b=1.0$, for which  $\Omega_{\rm m}=0.314^{+0.038}_{-0.035}$.

\begin{figure}
\vspace{0.5cm} \hbox{
\hspace{-0.5cm}\psfig{figure=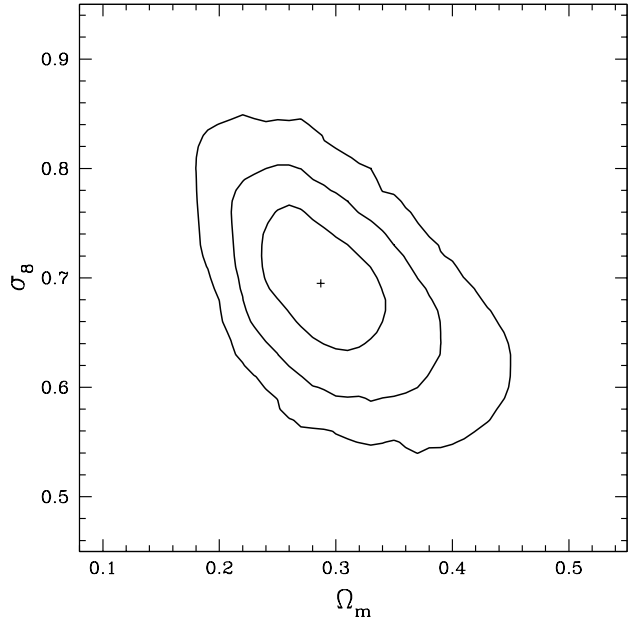,width=.49\textwidth,angle=0}
} \caption{The 68.3, 95.4 and 99.7 per cent confidence contours on
$\sigma_8$ and  $\Omega_{\rm m}$ from the combined analysis of the
BCS+REFLEX luminosity  function and Chandra $f_{\rm gas}(z)$ data. A
flat  $\Lambda$CDM cosmology is assumed.}\label{fig:contour2}
\end{figure}

We can now combine (by multiplying the relevant probability
densities) the  constraints on $\Omega_{\rm m}$ from
Fig.~\ref{fig:fgas_chi} with  the joint constraints on $\sigma_8$ and
$\Omega_{\rm m}$, shown in Fig.~\ref{fig:contour1}, to obtain our final
results on $\sigma_8$ and $\Omega_{\rm m}$. These are shown in
Fig.~\ref{fig:contour2}. Using the Gaussian prior on the bias
parameter, we obtain $\sigma_8=0.695\pm0.042$ and
$\Omega_{\rm m}=0.287\pm0.036$ (marginalized 68 per cent confidence
limits). With $b=1.0$ fixed, we obtain  $\sigma_8=0.683\pm0.041$ and
$\Omega_{\rm m}=0.309\pm0.035$.
 
\subsection{Systematic uncertainties and the effects of merger events}\label{section:systematics}

An important aspect of the present work is the reduced level of
systematic uncertainty with respect to most previous studies based on
the local abundance of galaxy clusters. In the first case, the
independent BCS and REFLEX studies have well determined selection
functions (Ebeling \etal 1998, 2000; B\"ohringer \etal 2002) and
provide precise, consistent results on the local X-ray luminosity
function. With the large size of the combined BCS-plus-REFLEX data set
(which covers two thirds  of the sky) we have been able to limit our
analysis to the most luminous clusters, with $L_{X,0.1-2.4}>
10^{45}\,h_{50}^{-2}$\ergps~(for an $\Omega_{\rm m}=0.3$,
$\Omega_\Lambda=0.7$, cosmology), for which the mass-luminosity
relation has been calibrated using  Chandra and ROSAT X-ray data and
weak lensing observations (Section~\ref{section:ml}). The complicating
effects of cooling and pre-heating are minimized for such massive
clusters and a simple power-law model provides a reasonable
description of the mass-luminosity relation (Fig.~\ref{fig:ml}),
albeit with some residual intrinsic scatter that is in part related to
the different dynamical histories of the clusters and residual
systematic effects in the  mass measurements
(Sections~\ref{section:dahle}, \ref{section:ml}; see also
below). Repeating  the analysis using only the BCS luminosity function
data, or only the REFLEX data, or using a higher luminosity cut in the
luminosity function (\eg $L_{X,0.1-2.4}> 1.3 \times
10^{45}\,h_{50}^{-2}$\ergps), leads to consistent results (although
with larger statistical uncertainties).  Consistent results on
cosmological parameters are also obtained if we replace the Evrard
\etal (2002) critical spherical overdensity mass function
(Section~\ref{section:theory}) with the Jenkins \etal (2001) mean
spherical overdensity (SO324) prescription, scaling the cluster
masses accordingly.

The dominant statistical and systematic uncertainties in the analysis
are associated with the mass-luminosity relation. However, care has
been taken to minimize the systematic uncertainties. In particular, we
have limited our study of the mass-luminosity relation to clusters
with precise mass measurements from Chandra or wide-field
gravitational lensing studies. The clusters studied with Chandra have
been selected from the RASS as the most luminous, dynamically relaxed
(in terms of their X-ray and optical morphologies) clusters known.
The relaxed natures of these systems means  that they are the clusters
for which X-ray mass measurements are most reliable. For Abell 963,
1835, 2390 and RXJ1347-1145, independent confirmation of the Chandra
mass measurements at $r_{200}$ is available from weak gravitational
lensing studies  (Dahle \etal 2002; Squires \etal 1996; Fischer \&
Tyson 1997; Allen \etal 2001a, 2002b).  A programme of weak lensing
measurements for the other clusters in our Chandra sample is underway
(Gray \etal,  in preparation).  In addition, consistent strong lensing
mass measurements are available for Abell 1835, 2390, RXJ1347.5-1145,
MS2137.3-2353 and PKS0745-191 (Schmidt \etal 2001, Allen \etal 2001a,
2002b, Schmidt \etal, in preparation).  The presence of significant
non-thermal pressure support on large spatial scales in these clusters
can therefore be excluded.

As well as increasing the size of our calibration sample, the
inclusion of the Dahle \etal (2002) weak lensing data is important  in
that it allows us to examine the distribution of less dynamically
relaxed systems in the mass-luminosity plane. Whereas X-ray data can
only be used to obtain precise mass measurements for dynamically
relaxed clusters (the X-ray mass measurements are based on the
assumption of  hydrostatic equilibrium in the X-ray gas), lensing mass
measurements are essentially independent of the dynamical state of the
gravitating  matter and so can be extended to systems undergoing
merger events.  The Dahle \etal (2002) sample includes clusters with a
range of dynamical states. Inspection of ROSAT and Chandra images
of these systems shows that many exhibit significant dynamical
activity, with Abell 141, 209, 520, 1351 and 1576 undergoing major
merger events. (Similar conclusions are drawn by Dahle \etal 2002 from
their optical data). The BCS and REFLEX samples, from which the X-ray
luminosity functions have been constructed
(Section~\ref{section:bcs}), include all clusters above the respective
X-ray flux limits, independent of their dynamical states. It is
therefore important that the mass-luminosity relation provides a fair
translation between luminosity and mass for all clusters included in
the samples. Although the 17 clusters in our mass-luminosity relation
do not represent a complete subsample of the BCS and REFLEX data sets,
inspection of Fig.~\ref{fig:ml} suggests that the Dahle \etal (2002)
clusters exhibit a similar mean mass per unit X-ray luminosity to the
relaxed systems studied with Chandra, albeit with two significant
outliers from the best-fit curve (see below) and, therefore, that the
application of the mass-luminosity relation determined here 
to the BCS and REFLEX samples is reasonable.

\begin{figure}
\vspace{0.5cm}
\hbox{
\hspace{-0.0cm}\psfig{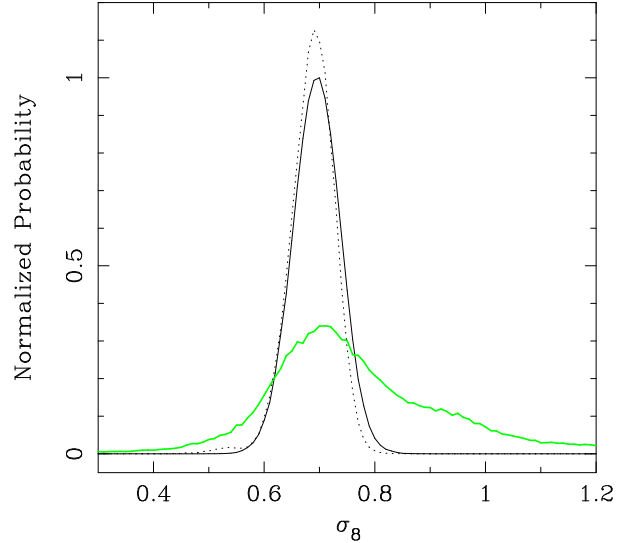}}
\caption{The normalized, marginalized probability distributions 
for $\sigma_8$ obtained from the Monte Carlo analysis using mass-luminosity 
relations constructed from the whole sample of
17 clusters (solid line), the subsample of 10 clusters 
with Chandra mass measurements (dotted line), and the 
subsample of 8 clusters with wide-field weak lensing mass 
measurements (grey line). See text for details.}\label{fig:marginal_s8}
\end{figure}

We have carried out a more rigorous test of the validity of the
mass-luminosity relation by repeating the determination of
cosmological parameters using other mass-luminosity relations
constructed from subsamples of the calibration data. In the first
case, a mass-luminosity relation based on the Chandra
mass measurements (10 clusters) was examined. In the second case, only
the weak lensing measurements were used (8 clusters).\footnote{For the
determination of cosmological parameters with the weak lensing
subsample, only the first two bins of the BCS and REFLEX luminosity
functions (Table~\ref{table:lumfunc}) were fitted, as the  Dahle \etal
(2002) clusters do not populate the highest luminosity bin.}   These
results were then compared with those obtained from the full sample (17
clusters, excluding the lensing result for Abell 963). The normalized,
marginalized probability distributions for $\sigma_8$ 
are shown in Fig.~\ref{fig:marginal_s8}. For the Chandra
subsample, we obtain $\sigma_8=0.69\pm0.04$. For the lensing
subsample, we find $\sigma_8=0.71^{+0.17}_{-0.12}$ (68 per cent confidence
limits). Both results are consistent with those obtained for the whole
sample, $\sigma_8=0.70\pm0.04$, although for the lensing subsample the
increased scatter in the mass-luminosity relation leads to larger error
bars. It appears, then, that the inclusion of Chandra mass measurements 
for dynamically relaxed clusters does not bias the best-fit result on
$\sigma_8$ significantly, but does reduce the formal statistical
uncertainties. Assessing whether this leads us to underestimate 
the true uncertainties in $\sigma_8$ will require more data and the 
ability to disentangle whether the scatter in the lensing 
results arises primarily from the effects of correlated 
substructure (or similar systematic problems in the lensing 
analysis), or the effects of dynamical activity on the X-ray gas 
(see below).

Ricker \&  Sarazin (2001), Ritchie \& Thomas (2002) and Randall,
Sarazin \& Ricker (2002) present simulations of the effects of mergers
on the X-ray properties of clusters for a variety of mass ratios,
impact parameters and central gas densities. These authors suggest
that major mergers can lead to significant short-term increases in the
luminosities of clusters during  the periods of closest approach,
which are then followed by dips in luminosity as the merging dark
matter cores move apart, before the cluster returns to
equilibrium. From Fig.~\ref{fig:ml} we see that Abell 209 appears to
have an unusually high X-ray luminosity/mass ratio, which may have
been boosted by the ongoing merger activity in this cluster.  (We
note, however, that MS2137.3-2353, which is a highly relaxed cluster
with no obvious merger activity and a sharp central density peak,
also appears to have a high luminosity/mass ratio and lies below the
best-fitting curve. In this case, the offset may indicate a relatively 
early formation epoch for the cluster.) In contrast Abell 1351, which is
also undergoing a major merger event, has an unusually low X-ray
luminosity/mass ratio and lies above the best-fit curve in
Fig.~\ref{fig:ml}. Detailed simulations  and further observations are
required to improve our understanding of  the effects of mergers on
the X-ray properties of clusters.  However, the indications from the
present study are that the effects of mergers on the mean mass-luminosity 
relation for the most luminous clusters are relatively small.

Chandra measurements of  the X-ray gas mass fraction, $f_{\rm gas}$,
in  dynamically relaxed clusters provide one of  the most simple and
robust methods by which to measure $\Omega_{\rm m}$ (see \eg Allen
\etal 2002a and references  therein). The largest systematic
uncertainties in the analysis lie in how well the measured baryonic
mass fractions in the clusters approximate the universal mean.
Lowering the bias factor, $b$, by $\sim 10$ per cent would cause the
best-fitting value of $\Omega_{\rm m}$ to rise by a similar amount.
Similarly, although the $f_{\rm gas}$ profiles in
Fig.~\ref{fig:fgas_r} appear to have converged, or be close to
converging, at the outer measurement radii, any rise in the $f_{\rm
gas}$ values beyond these points would cause a corresponding reduction
in  $\Omega_{\rm m}$. Any change in the best-fitting value of
$\Omega_{\rm m}$ would then  lead to a change in $\sigma_8$, as shown
in Fig.~\ref{fig:contour1}.

Finally, we note that our Monte Carlo analysis (Section 4.1) takes
account of the  intrinsic scatter and uncertainties in the
normalization and slope  of the mass-luminosity  relation. (The
scatter is modelled as a log-normal distribution about the
best-fitting curve, which provides a good description of the measured
offsets.) Neglecting such effects would cause to the best-fitting
value of $\sigma_8$ to rise slightly, and significantly reduce the
formal statistical uncertainties on the measured parameters.

\section{Comparison with other results}\label{section:compare}

\subsection{Other local cluster abundance studies}

\begin{figure}
\vspace{0.5cm}
\hbox{
\hspace{0.2cm}\psfig{figure=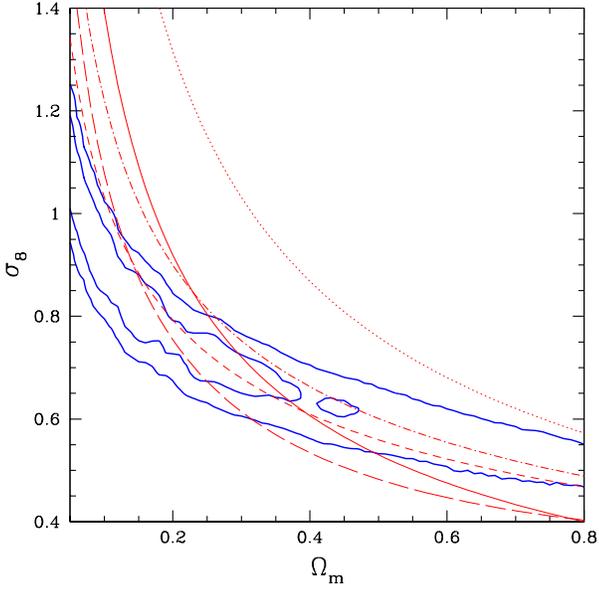,width=.46\textwidth,angle=0} }
\caption{The 68.3 and 95.4 per cent confidence contours on $\sigma_8$ as a
function  of $\Omega_{\rm m}$ determined from the present study
(thick, solid curves; as in Fig.~\ref{fig:contour1}) together with the
best-fit results of Seljak (2002: dot-dashed curve), Pierpaoli \etal (2001:
dotted curve), Reiprich \& B\"ohringer (2001: short-dashed curve),
Viana \etal (2002: long-dashed curve) and  Bahcall \etal (2002: thin,
solid curve). See text for details. A flat $\Lambda$CDM cosmology  is
assumed.}\label{fig:seljak}
\end{figure}

Fig.~\ref{fig:seljak} shows a comparison of the results on $\sigma_8$
as a function of $\Omega_{\rm m}$ from the present study (thick, solid
curves: as in Fig.~\ref{fig:contour1}), with the findings from five
other recent studies based on the observed local number density of
galaxy clusters. The dot-dashed curve in Fig.~\ref{fig:seljak} shows
the result of  Seljak (2002), $\sigma_8 = (0.44\pm0.04) \,\Omega_{\rm
m}^{-0.44} (\Gamma/0.2)^{0.08}$, using the observed local X-ray
temperature function of rich clusters (Pierpaoli,  Scott \& White
2002) and a mass-temperature relation determined from  ASCA and ROSAT
observations (Finoguenov \etal 2001).  The result of Seljak is consistent
with ours for $\Omega_{\rm m} \sim 0.3$, although the
present study leads  to tighter constraints. The result of Pierpaoli
\etal (2001), $\sigma_8 = (0.50\pm0.04)\, \Omega_{\rm m}^{-0.60}$
(dotted curve), based on the same local temperature function data but
normalized by a theoretical mass-temperature relation, lies
significantly above ours. The result of Reiprich \& B\"ohringer
(2001), $\sigma_8 = 0.43\, \Omega_{\rm m}^{-0.38}$ (short dashes),
based on ASCA and ROSAT observations of RASS selected clusters, is
in good agreement with the present study for $\Omega_{\rm m} \sim 0.3$.  The
result of  Viana, Nichol  \& Liddle (2002), $\sigma_8 = 0.38\,
\Omega_{\rm m}^{-0.48+0.27\,\Omega_{\rm m}}$ (long dashes), using the
local REFLEX X-ray luminosity function and a mass-luminosity relation
determined from ROSAT X-ray observations and a stacked weak lensing
analysis of relatively low-mass clusters identified in Sloan Digitized
Sky Survey (SDSS) commissioning data, is consistent with the present study 
for $\Omega_{\rm m} \sim 0.3$. The result of Bahcall
\etal (2002), $\sigma_8 = 0.35\, \Omega_{\rm m}^{-0.60}$, obtained by
combining the number density of optically-selected, relatively
low-mass clusters observed in SDSS commissioning data with a
mass-optical richness  correlation, is in good agreement with this
work for $\Omega_{\rm m} \sim 0.3$. Finally, our results are in  good
agreement with the recent findings of Schuecker \etal (2003;  not
plotted), $\sigma_8=0.711^{+0.039}_{-0.031}$,  $\Omega_{\rm
m}=0.341^{+0.031}_{-0.029}$, from a combined analysis of  the X-ray
luminosity function and large-scale clustering in the  REFLEX sample.

\subsection{Evolution in the X-ray luminosity and temperature functions}\label{section:evolution}

Borgani \etal (2001) present constraints on $\sigma_8$ and
$\Omega_{\rm m}$ determined from an analysis of evolution in the X-ray
luminosity function of clusters in the ROSAT Deep Cluster Survey,
which spans the redshift range $z \approxlt 1.3$. Their results of
$\sigma_8=0.66^{+0.06}_{-0.05}$ and  $\Omega_{\rm
m}=0.35^{+0.13}_{-0.10}$ for a flat $\Lambda$CDM cosmology are
consistent  with those reported here.

Donahue \& Voit (1999) report results from an analysis of evolution in
the temperature function of clusters within $z \approxlt 0.8$, using
the low-redshift cluster sample of Markevitch (1998) and a high
redshift sample identified from the Einstein Observatory  Extended
Medium Sensitivity Survey (EMSS). Their results, for an  assumed flat
$\Lambda$CDM cosmology, of $\sigma_8=0.73^{+0.03}_{-0.05}$  and
$\Omega_{\rm m}=0.27\pm0.10$ are in good agreement with ours.  Eke
\etal (1998b) obtain  $\sigma_8=0.75\pm0.15$ and $\Omega_{\rm
m}=0.36\pm0.25$  (flat $\Lambda$CDM) from an analysis of the
temperature function  within $z\sim0.4$, which is consistent with the
present work. Our results are marginally consistent with the values
of $\sigma_8=0.72\pm0.10$ and $\Omega_{\rm m}=0.49\pm0.12$ reported by
Henry (2000) from an analysis combining  the local
temperature function of clusters with the properties of  EMSS clusters
within $z<0.6$.

\subsection{Cosmic shear measurements}

\begin{figure}
\vspace{0.5cm} \hbox{
\hspace{0.2cm}\psfig{figure=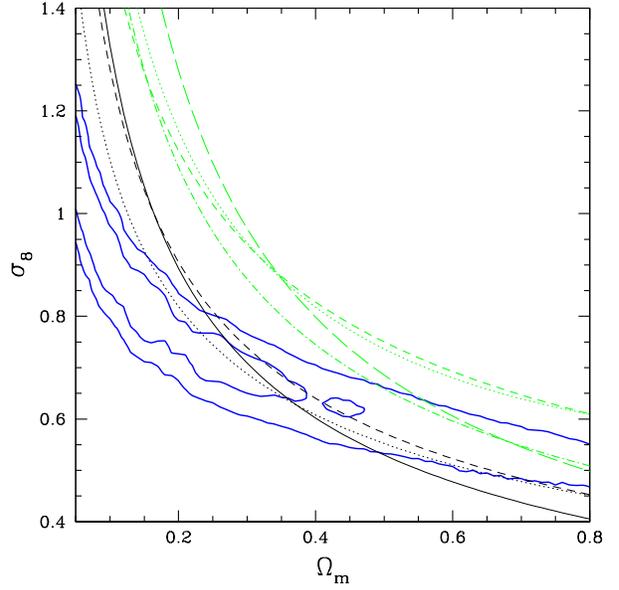,width=.46\textwidth,angle=0}
}
\caption{The 68.3 and 95.4 per cent confidence contours on $\sigma_8$ as a
function of $\Omega_{\rm m}$ from the present study (dark, solid
curves;  as in Fig.~\ref{fig:contour1}) together with the best-fit
results from the cosmic shear studies of  Van Waerbeke \etal (2002:
upper, dotted curve), Refregier \etal (2002:  upper, short-dashed
curve), Bacon \etal (2002: long-dashed curve), Hoekstra \etal (2002:
dot-dashed curve), Hamana \etal (2002: lower, dotted curve), Brown
\etal (2002: lower, dashed curve) and Jarvis \etal (2002: solid
curve),  A flat $\Lambda$CDM cosmology is assumed.}\label{fig:shear}
\end{figure}

Fig.~\ref{fig:shear} shows a comparison of the results on $\sigma_8$
as a function of $\Omega_{\rm m}$ determined from the present study
(dark, solid curves: as in Fig.~\ref{fig:contour1}) together with the
findings from seven recent studies based on measurements of weak
gravitational lensing due to large scale structure (cosmic shear). The
upper dotted curve in Fig.~\ref{fig:shear} shows the result of Van
Waerbeke \etal (2002),  $\sigma_8 = (0.57\pm0.04)\, \Omega_{\rm
m}^{(0.24\pm0.18)\Omega_{\rm m}-0.49}$.  The upper short-dashed curve
shows the result of  Refregier, Rhodes \& Groth (2002), $\sigma_8
=(0.55\pm0.08)\, \Omega_{\rm m}^{-0.44}$. The long-dashed curve shows
the result  of Bacon \etal (2002), $\sigma_8 = (0.43\pm0.06)\,
\Omega_{\rm m}^{-0.68}$. The dot-dashed curve shows the result of
Hoekstra, Yee \& Gladders (2002),  $\sigma_8
=(0.45\pm0.05)\,\Omega_{\rm m}^{-0.55}$.  The results on $\sigma_8$
from these four studies lie $20-35$ per cent above the present work
for $\Omega_{\rm m} \sim 0.3$. (Note that in order to obtain agreement
with these results, the mean mass per unit X-ray luminosity for the
clusters included in the present study  would need to be raised by a
factor $\sim 2.5$, well beyond the systematic uncertainties,  which
are $\approxlt 20$ per cent.)

The lower three curves  in Fig.~\ref{fig:shear} show the results from
the three most recent cosmic shear studies, which have appeared on
preprint servers after this paper was submitted.  The lower dotted
curve shows the result of Hamana \etal (2002), $\sigma_8
=(0.41\pm0.09)\,\Omega_{\rm m}^{-0.43}$. The lower dashed curve shows
the result of Brown \etal (2002), $\sigma_8
=(0.41\pm0.05)\,\Omega_{\rm m}^{-0.50}$. Finally, the solid curve
shows the result of Jarvis \etal (2002),  $\sigma_8
=(0.36\pm0.04)\,\Omega_{\rm m}^{-0.57}$. Importantly, the Brown \etal
(2002) and  Jarvis \etal (2002) studies include improved estimates of
the redshifts of the lensed, background sources. (A flat $\Lambda$CDM
cosmology is assumed  in all cases.)  The results from the three most
recent cosmic shear studies are in good agreement with the present
work.

\subsection{Cosmic microwave background anisotropies and the 2dF galaxy 
redshift survey}

\begin{figure}
\vspace{0.5cm} \hbox{
\hspace{-0.4cm}\psfig{figure=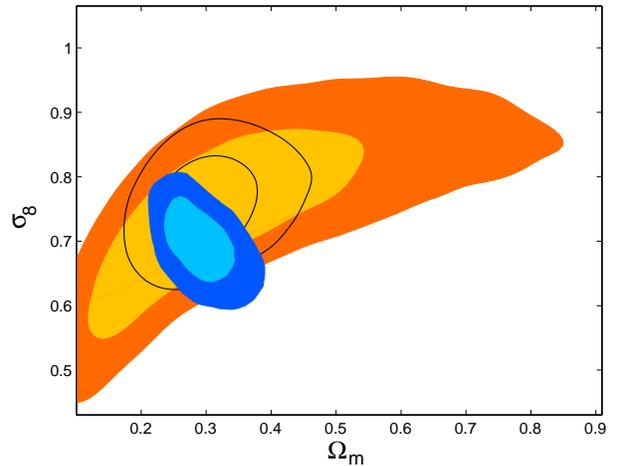,width=.51\textwidth,angle=0}
} \caption{The 68.3 and 95.4 per cent confidence contours on $\sigma_8$
and $\Omega_{\rm m}$ from the analysis of CMB (outer, filled contours)
and CMB+2dF data (solid lines). An optical depth to  reionization,
$\tau=0.04$ is assumed. We have marginalized over  all other
parameters ($h$, $\Omega_{\rm b}h^2$ and $n$).  The inner, filled
contours show the  results from the present study (as in
Fig.~\ref{fig:contour2}), including  Gaussian priors of
$h=0.72\pm0.08$, $b=0.93\pm0.05$ and  $\Omega_{\rm b}h^{2} = 0.0205\pm0.0018$.  A flat
$\Lambda$CDM cosmology is assumed in all cases.}\label{fig:sarah}
\end{figure}

Fig.~\ref{fig:sarah} shows a comparison of the 68.3 and 95.4 per cent
confidence constraints on $\sigma_8$ and $\Omega_{\rm m}$ obtained
from the present study (inner, shaded contours), including the
Chandra $f_{\rm gas}(z)$ data and Gaussian priors on the bias
parameter ($b=0.93\pm0.05$), Hubble
constant ($h=0.72\pm0.08$) and $\Omega_{\rm b}$ ($\Omega_{\rm b}h^{2}
= 0.0205\pm0.0018$), with the results from analyses of cosmic
microwave background (CMB) anisotropies and 2dF galaxy redshift survey
data. For this comparison, we have used the published  Markov Chain
Monte Carlo samples of Lewis \& Bridle (2002). The CMB data  consist
of a combination of COBE (Bennett \etal 1996), Boomerang  (Netterfield
\etal 2002),  Maxima (Hanany \etal 2000), DASI (Halverson \etal 2002),
Cosmic Background Imager (Pearson \etal 2002) and Very Small Array
(Scott \etal 2002) data.  The 2dF galaxy redshift survey constraints
are from Percival \etal (2002).   The results obtained from the CMB
data alone, using the 6 parameter model of Lewis \& Bridle (2002) and
fixing the optical depth to  reionization $\tau=0.04$ and
marginalizing over $h$, $\Omega_{\rm b}h^2$ and $n$, are shown as the
outer, shaded contours. The results obtained from  the
combined CMB+2dF data set, marginalizing over the same parameters, are
shown as solid lines.  A flat $\Lambda$CDM cosmology is assumed
throughout. We see that the results from all three data sets are
consistent at the 68 per cent confidence level. Our results are also
in good agreement with the independent analyses of CMB+2dF data
reported by Lahav \etal (2002), Percival \etal (2002) and Melchiorri
\& Silk  (2002), the results from  CMB data and Type Ia supernovae
presented by  Jaffe \etal (2001), and the analysis of the IRAS Point
Source Catalogue Redshift Survey  presented by Plionis \& Basilakos
(2001).

\section{Implications for other work}\label{section:implications}

The constraints on cosmological parameters reported here
($\sigma_8=0.695\pm0.042$ and $\Omega_{\rm m}=0.287\pm0.036$ for an
assumed flat $\Lambda$CDM cosmology) are consistent with, though
(in most cases) tighter than, those obtained from a number of other, recent 
studies based on the observed X-ray temperature and luminosity functions of 
galaxy clusters. (Schuecker \etal 2003 report results with  
similar precision to those presented here).  Our results are also
consistent with current findings from studies of anisotropies in the
CMB, the distribution  of galaxies in the 2dF galaxy redshift survey,
the properties of type Ia supernovae, early results on large scale
structure from the SDSS, and the most recent results from studies
of cosmic shear.

Our results have a number of implications for other  cosmological
work.  Firstly, the agreement between the cluster, CMB and 2dF
results in Fig.~\ref{fig:sarah} suggests that the optical depth to
reionization is not large ($\tau \approxlt 0.2$; this issue will be
examined in more detail in future work). Secondly, our result on
$\sigma_8$ implies that the possible excess power detected at high
multipoles ($l\sim 2000-3000$) in the CMB anisotropy  power spectrum
with the Cosmic Background Imager (Mason \etal 2002;  Bond \etal 2002)
is unlikely to be due to the Sunyaev-Zeldovich (SZ)  effect.  Our
results also have important implications for  future X-ray and SZ
clusters surveys, since the number of clusters  detected at  high
redshifts with high X-ray luminosities, temperatures and SZ fluxes
will be much lower in a $\sigma_8 \sim 0.7$ universe than predicted
by simulations with $\sigma_8 \sim 1$.

Finally, we suggest that given the precision of the constraints on
$\Omega_{\rm m}$ available from current Chandra $f_{\rm gas}(z)$ data
for relaxed clusters, and the  relatively small systematic
uncertainties  involved, the $f_{\rm gas}(z)$ data should be
considered as a powerful probe in future cosmological work,
complementary to measurements of redshift evolution in the X-ray
temperature and luminosity functions of galaxy clusters.

\section*{Acknowledgements}

We thank George Efstathiou, Sarah Bridle and Jerry Ostriker  for a
number of helpful discussions and suggestions. We also thank Sarah
Bridle for providing the CMB and 2DF results shown in
Fig.~\ref{fig:sarah}. We thank Alastair Edge  and Carolin Crawford for
their efforts in compiling the  extended BCS sample, and Ofer Lahav,
John Peacock and Andrew Liddle for comments on the manuscript. SWA and
ACF acknowledge the support of the Royal Society. HE acknowledges
financial support from NASA grant NAG 5-8253 and CXO grant GO1-2132X.

\end{document}